# Electronics design and development of Near-Infrared Imager, Spectrometer and Polarimeter.

Deekshya Roy Sarkar*,[1], Amish B. Shah[1], Alka Singh[1], Pitamber Singh Patwal[1], Prashanth Kumar Kasarla[1], Archita Rai[1,2], Prachi Vinod Prajapati[1], Hitesh Kumar L. Adalja[1], Satya N. Mathur[1], Sachindra Naik[1], Shashikiran Ganesh[1], Kiran S. Baliyan[1]
deekshya@prl.res.in, abshah@prl.res.in, alkasingh@prl.res.in
1. Physical Research Laboratory, Ahmedabad, Gujarat, INDIA
2. Indian Institute of Technology, Gandhinagar, Gujarat, INDIA

## ABSTRACT

NISP, a multifaceted near-infrared instrument for the upcoming 2.5m IR telescope at MIRO Gurushikhar, Mount Abu, Rajasthan, India is being developed at PRL, Ahmedabad. NISP will have wide (FOV = 10' x 10') field imaging, moderate (R=3000) spectroscopy and imaging polarimetry operating modes. It is designed based on 0.8 to 2.5µm sensitive, 2048 X 2048 HgCdTe (MCT) array detector from Teledyne. Optical, Mechanical and Electronics subsystems are being designed and developed in-house at PRL. HAWAII-2RG (H2RG) detector will be mounted along with controlling SIDECAR ASIC inside LN2 filled cryogenic cooled Dewar. FPGA based controller for H2RG and ASIC will be mounted outside the Dewar at room temperature. Smart stepper motors will facilitate motion of filter wheels and optical components to realize different operating modes. Detector and ASIC temperatures are servo controlled using Lakeshore's Temperature Controller (TC) 336. Also, several cryogenic temperatures will be monitored by TC for health checking of the instrument. Detector, Motion and Temperature controllers onboard telescope will be interfaced to USB Hub and fiber-optic trans-receiver. Remote Host computer interface to remote end trans-receiver will be equipped with in-house developed GUI software to control all functionalities of NISP. Design and development aspects of NISP Electronics will be presented in this conference.

## 1. INTRODUCTION

This paper describes the development and working of Inhouse PRL make controller alongside its capabilities, SIDECAR ASIC RT and Cryo kit functional testing and its results, different software development work being carried out and the progress so far.

Because of its reliability and good performance, the H2RG focal plane arrays (FPA's) have been used in various observatories in India and abroad for its ground-based instrumentation. Based on our requirement, we are using the H2RG with 2.5µm cutoff substrate removed MCT detector which will operate at approximate 77K (LN$_2$). When completed, NISP will become a multi facility instrument for 2.5m telescope at Mount Abu InfraRed Observatory, Gurushikhar. There are two ways in which the Teledyne make H2RG detector will be controlled SIDECAR ASIC and Inhouse PRL make Controller (IPC). Teledyne has provided with two SIDECAR ASIC Kit configurations Room Temperature (RT) Kit and Cryogenic Kit. Both have been checked for functional testing. The SIDEACAR ASIC is Teledyne's custom-made ASIC which has its own micro controller and needs an external interface board according to its architecture for driving it and to establish communication with PC.

Besides SIDECAR ASIC, The Inhouse PRL make controller along with its acquisition software has been designed because it is cost effective and can be customized according to the user needs. It has been tested with NICMOS detector in 1 and 4 output modes. Various independent scripts of different modules such as temperature control, motion control and others have been programmed and tested individually. Integration of all these modules with main software is in pipeline.

## 2. INHOUSE CONTROLLER

### 2.1 Hardware

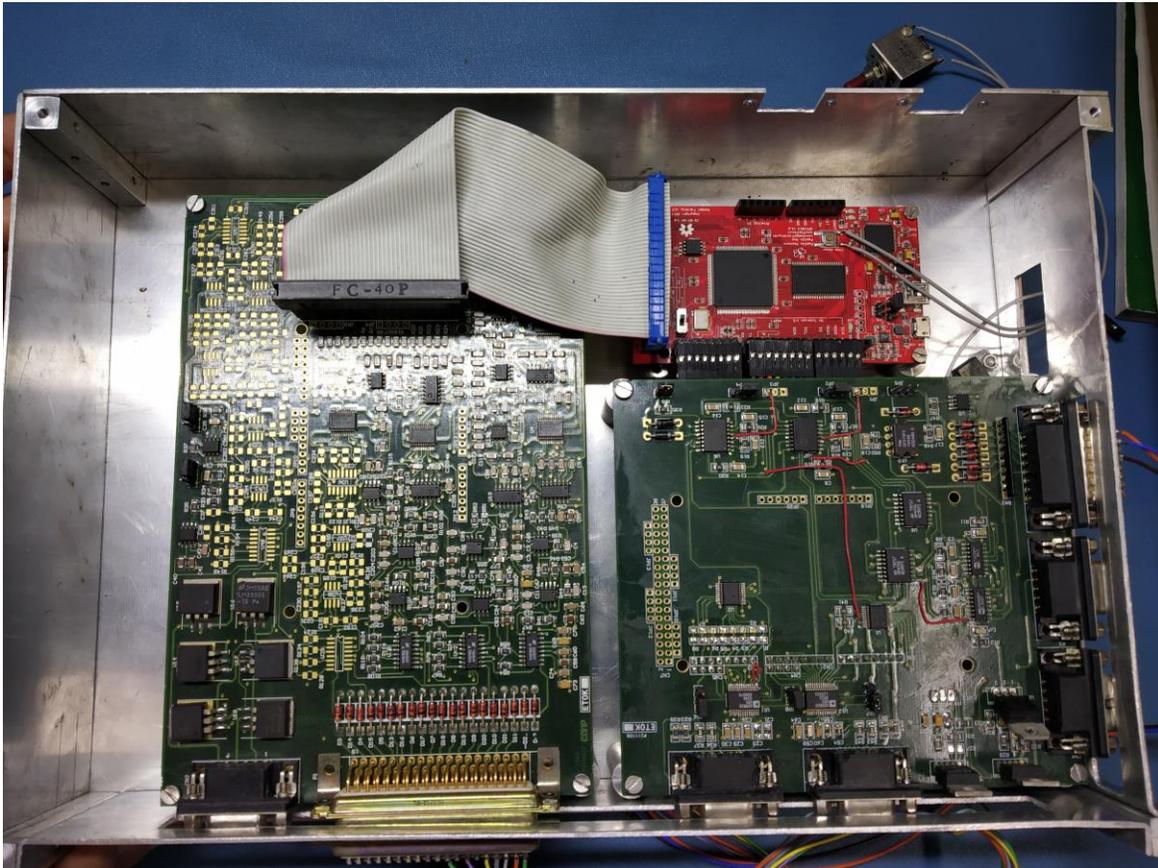

Figure 1. Open source FPGA development board (red color) based PRL NISP navigator is developed. Clock generator board (left side) and Bias – ADCs board (right side) are developed using discrete electronics.

The main motivation to develop our own controller was that the Teledyne provided SAM card for SIDECAR ASIC control has many proprietary modules which are obviously undocumented and is hard to interface with astronomical packages. As shown in Figure1, the Inhouse PRL make controller is based around Xilinx Spartan 6 FPGA in Papilio Duo board. It has two cards clock board and ADC & Bias board. All the necessary voltages required for operation of a detector can be generated by the bias board. Both cards have on board linear regulators through which power is given to its every component. But as of now we have the liberty of 1 and 4 no of outputs which later on can be expanded to 32 output mode.

### 2.2 Integrated testing with NICMOS detector

Near Infrared Camera and Multi-Object Spectrometer (NICMOS) is based on HgCdTe detector array and is sensitive from 1 – 2.5 µm. It was in use at PRL's 1.2m telescope at Mt Abu Gurushikhar from 1997 to late 2000 after which other new instruments took over. Therefore, to check our controller for detector control we used NICMOS's 256 X 256 array detector. A pre amplifier board was designed to interface our electronics with NICMOS output signals and all the clock signals were generated by the FPGA.

The most peculiar thing about IPC is it can be programmed to control variety of CCD and CMOS detectors. A NISPSOF has been developed in C# for windows platform. It will have integrated package for image acquisition, image analysis, detector temperature control and optics temperature monitoring, operating modes selection and motion control, data communication and interface for Remote Telescope (RTS2) automated observatory open source software.

## 3. SIDECAR ASIC ROOM TEMPERATURE KIT SETUP

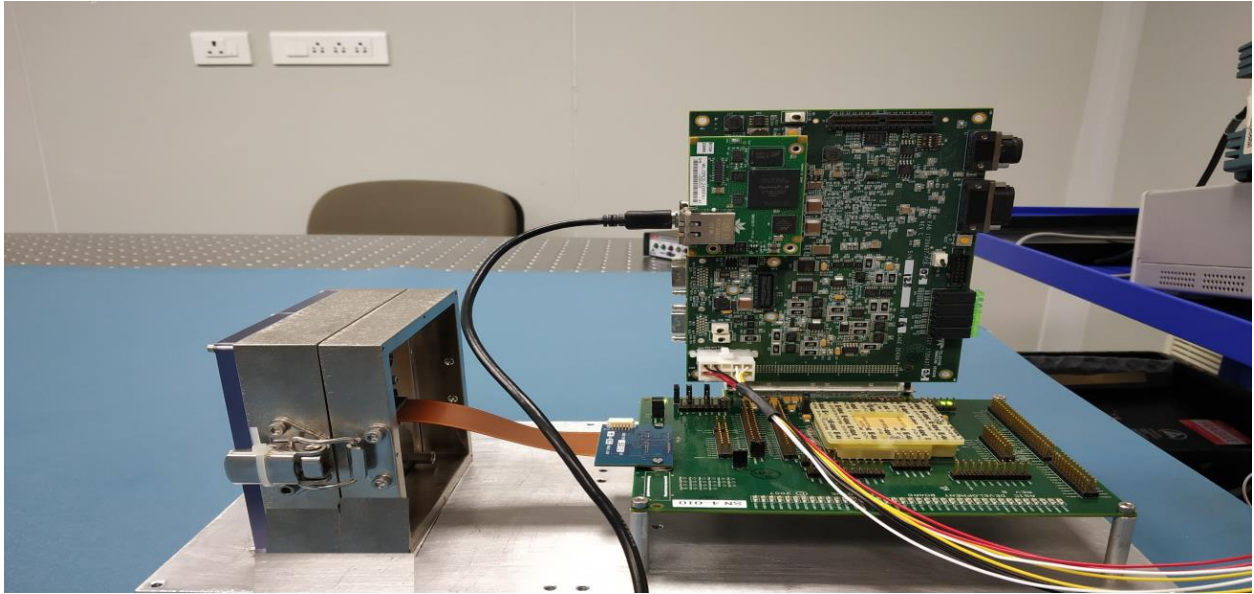

Figure 2. ROIC multiplexer (left side) , SIDECAR warm development board (right side) and SAM board (upwards)

### 3.1 Test Setup

We have used the Teledyne make Sidecar ASIC, 337-LGA warm development board along with Sidecar Acquisition Module (SAM) board. The entire assembly is been assembled with the bare ROIC/ H2RG multiplexer with the 1.5" flex cable. The ROIC is a silicon multiplexer layer of H2RG detector having identical electrical and mechanical interface of H2RG but without the light sensitive layer. Figure2 shows the assembled setup. Aluminum mounts house both the ROIC and SIDECAR warm board. In this setup, the connection to the SAM board from SIDECAR ASIC board is done vertically 90° through Molex connector provided by Teledyne. To avoid any interference, separate wires of analog and digital grounds to SAM board were drawn and brought to a common ground. Clean power of 5.5 V was given to SAM board through external power supply by disabling jumper in SAM board, USB 2.0 was used only for data and command transfer.

### 3.2 Software

Teledyne has provided two software packages called Interactive Development Environment (IDE) and HxRG socket server also known as Interactive Data Language (IDL) software. All the Sidecar ASIC features can be accessed by using IDE. It provides USB2.0 communication interface to communicate with ASIC hardware. The microcontroller's firmware can be reprogrammed, altered and debugged using IDE. All necessary voltage and current biases are enabled by IDE.

The HxRG socket server is implemented in IDL and is mainly a readout software to acquire image and store in PC in fits format.

We operated the IDE & IDL in a manner suggested by Teledyne. Although IDE is capable of fully configuring the SIDECAR ASIC but when IDL is switched on, the required files for ASIC and detector can be loaded from IDL automatically. Slow mode and fast mode operability function files can also be accessed through IDL. Some programmable exposure parameters of HxRG and SIDECAR like multiplexer type (e.g. H1RG, H2RG, H4RG), no. of detector outputs (e.g. 1,4,32), detector readout mode, SIDECAR PreAmp gain, acquisition mode (e.g. up the ramp, Fowler) can be configured using IDL.

## 4. ACQUISITION SOFTWARE DESIGN

The NISP control software named "NISP Navigator" is being developed for MS Windows environment. At the moment, image acquisition software, motion control and temperature monitoring segments have been developed and are working properly, all other segments are still under development and testing phase. It has been designed using python programming language. Figure3 shows the main building blocks of the NISP control software:

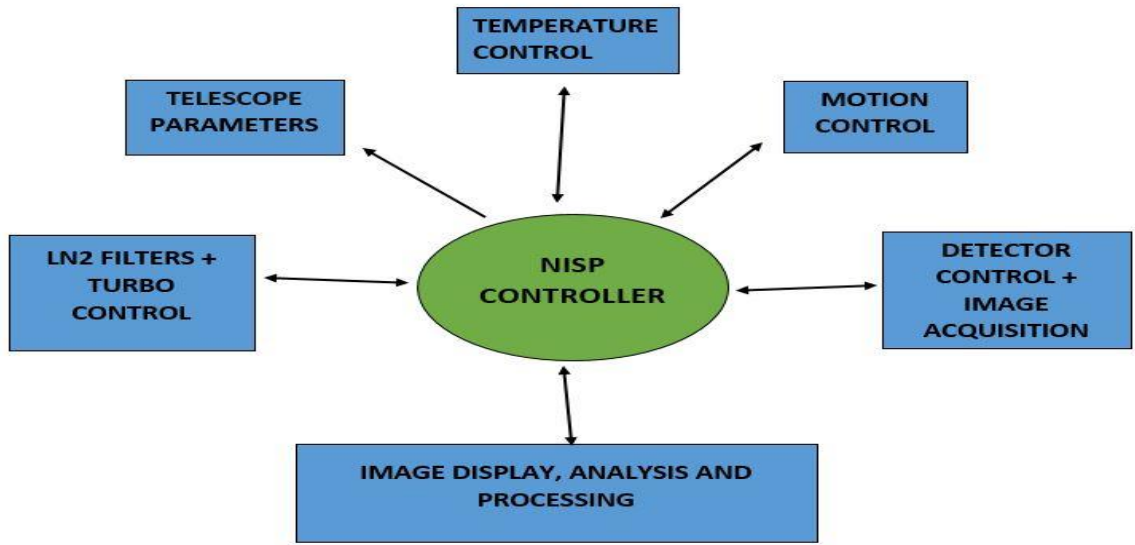

Figure 3. Main architecture blocks for the NISP control software.

- Monitoring all temperature sensors inside the cryostat is very crucial and is a continuous process. This is performed by Silicon Diodes (DT-670-SD) and Temperature Controller 336. The diodes is connected at several points inside the dewar: two near optics, one at SIDECAR mount plate and one at detector base plate mounting. Additionally, the SIDECAR ASIC and H2RG detector comes with temperature monitoring capabilities and ASIC with heating points. These diodes show reliable use from 1.4K to 500K and were chosen because it is very accurate at 77K which is our desire point of Liquid Nitrogen (LN2) temperature.

- Motion control module is capable to control eight filter wheels one at time. Three filter wheels are used to accommodate all required optical components. Each filter wheel is controlled by its dedicated Smart Stepper motor. Each filter wheel has six circular slots to accommodate 2" diameter filter/component. Each filter has absolute ID sensor. Using GUI any filter from selected filter wheel can be bring into optical path for desire operating mode.

- Detector control and Acquisition module will be either Teledyne supplied IDE software or in house developed image acquisition – NISPSOFT software. It will control H2RG and cryogenic ASIC through either SAM or PRL FPGA board. All initialization and configuration will be carried out by this module. Images will be acquired as per selected technique – Double Correlated Sampling, Up-The-Ramp, Fowler frames -. Acquired images will be converted into standard FITS format.

- Image display, Analysis and processing software will be developed in Python. Teledyne supplied IDL GUI also will be integrated with this module. It will do instrument correction, image filtering and image analysis. Also, as per scientist requirement some customize processing also will be provided by this module for on-line image processing.

- Telescope Parameters module will read necessary telescope housekeeping information such as RA, DEC, Object, UT etc. to be embedded into FITS file.

- We are also aiming to control LN2 filling and dewar evacuation automatically. LN2 filler and turbo control module will perform this task as per programmed schedule. Development of this module is in TBD state.

## 5. CONCLUSION

- Electronics design and development plan were presented to Technical Evaluation committee (TEC), PRL before being carried out.

- H2RG specifications and its performance test report were evaluated in consultation with TEC.

- Indigenous NISP controller is developed and tested with NICMOS3 detector. This has boost confidence to handle detector & electronics.

- RT SIDECAR setup being used for all laboratory tests and characterization.

- Motion control electronics integrated with filter wheels and GUI tested.

- Temperature control electronics integrated with detector mounting inside cryogenic dewar and being characterized.